\documentclass[final]{ias3}

\usepackage{graphicx} 
\usepackage{multirow}
\usepackage{array} 
\usepackage{amssymb}

\usepackage{hyperref} 
\begin{document}

%\markboth{Pramana class file for \LaTeX 2e}{Xerxes Yu, et. al.}
\markboth{Stellar Dynamics and Black Holes}{David Merritt}

\title{Stellar Dynamics and Black Holes}
\author[dm]{David Merritt}
\address[dm]{Rochester Institute of Technology, Rochester, NY, USA}

 \maketitle

\section{Introduction}

Chandrasekhar's most enduring contribution to stellar dynamics is
probably dynamical friction. 
The history of that discovery is fascinating and well known \cite{Padmanabhan1996}.
The first (1942) edition of {\it Principles of Stellar Dynamics}
contained a chapter entitled ``The Time of Relaxation of a Stellar System,''
in which Chandrasekhar presented his ``completely rigorous'' calculation
of the rate at which a star is accelerated by encounters
with other stars.
The result was expressed in terms of $\sum\Delta E^2$, 
the accumulated squared kinetic energy change after some time $\Delta t$.
From this, Chandrasekhar defined the relaxation time as
\begin{equation}
\frac{\sum\Delta E^2}{E^2} = \frac{\Delta t}{T_E}\nonumber
\end{equation}
and showed that $T_E$ was very long, of order $10^{12}$ yr, for a system
like the Milky Way.
He then defined a second relaxation time , $T_D$,
in terms of the angular deflection of an orbit and showed that it was
equivalent to $T_E$.

It was apparently only after publication of his monograph that Chandrasekhar
recognized another important consequence of gravitational encounters.
In the absence of a decelerating force, repeated encounters would 
increase the kinetic energy of a star without limit.
In a fluid medium, Stokes' law states that there is a frictional force,
proportional to the velocity, that competes with the acceleration.
In a paper published in 1943 \cite{Chandrasekhar1943}, Chandrasekhar 
recognized the necessity for such a force in the stellar dynamical
case, and showed that the assumption of a Maxwellian velocity distribution
implied a simple relation between the frictional force
and his previously-derived coefficient of diffusion. 
He then derived the coefficient of dynamical friction by summing the
velocity changes experienced by a test body, in the direction of its motion, 
as field stars were deflected around it.

Chandrasekhar was clearly proud of the fact that Brownian 
motion in stellar systems
could be treated in such an exact way, compared with
the ``intuitive character of the assumptions'' in the case of fluids \cite{Chandrasekhar1949}.
He wrote:
\begin{quotation}
The discussion of the physical foundations of the theory of Brownian motion 
[in fluids] in the
preceding sections has disclosed certain inherent limitations in the theory.
The limitations are nowhere more serious than in the circumstance that the
coefficients $q$ and $\eta$ [the coefficients of diffusion and friction]
are not derived from a microscopic analysis of the individual encounters.
It is therefore of interest that stellar dynamics provides a case of Brownian
motion in which all phases of the problem can be explcitly analyzed.
\end{quotation}

In this lecture, I briefly review Chandrasekhar's derivation of the
dynamical friction cofficient, then discuss its application in a few cases
that are relevant to the nuclei of galaxies containing massive 
black holes.
I conclude by discussing recent work that extends Chandrasekhar's ideas 
to star clusters in which the effects of relativity can not be ignored.

\section{Dynamical Friction}

In deriving the velocity diffusion coefficients in a stellar
system, Chandrasekhar
assumed that the instantaneous forces acting on a test star could be separated into
two components: a component due to the smoothed-out distribution of matter,
and a component arising from chance stellar encounters.
He assumed that the encounters were independent, and computed  their
net effect by integrating the momentum changes
of single encounters over the variables defining
the relative orbit of the binary (test star, field star) system.
The smoothed-out component was assumed to be infinite and homogeneous,
implying a constant velocity for the test mass in the absence of encounters.
Aside from enforcing istropy in velocity space, no restrictions were placed
on $f(\boldsymbol{u})$, the distribution of field star velocities at infinity.

Integrating over all impact parameters $p$, 
the change per unit interval of time of the test star's velocity, 
in a direction parallel to the initial relative motion, is then given by
\begin{equation}
\overline{(\Delta v_{\parallel})} = -{2\pi G^2m_f\left(m_f+m\right) n\over V^2}\ln\left(1+p^2_\mathrm{max}/p_0^2\right).
\label{Equation:coef1}
\end{equation}
Here $m$ and $m_f$ are the masses of the test and field star respectively,
$n$ is the field star density,
 $\boldsymbol{V}=\boldsymbol{v}-\boldsymbol{u}$ is the relative velocity
of test and field star before the encounter,
and
\begin{equation}
p_0 = \frac{G\left(m_f+m\right)}{V^2}\nonumber
\end{equation}
is the impact parameter corresponding to an angular deflection of $\pi/2$.
The final step is the integration over field star velocities.
Since equation~(\ref{Equation:coef1}) gives the velocity change in the direction
of the initial {\it relative} motion, we must multiply it
by 
\begin{equation}
\frac{\boldsymbol{V}\cdot\boldsymbol{v}}{Vv} = 
\frac{v-u_{x}}{V}\nonumber
\end{equation}
to convert it into a velocity change in the direction of the {\it test star}'s
motion, assumed here to be along the $x$ axis.
%Let$f(\boldsymbol{u})\boldsymbol{du}$ be the number density of
%field stars in velocity increment $\boldsymbol{u}, \boldsymbol{u}+\boldsymbol{du}$,
%normalized to unit total number.
The dynamical friction coefficient is then
\begin{eqnarray}
\langle\Delta v_{\parallel}\rangle & = & \int f(\boldsymbol{u})\, \overline{(\Delta v_{\parallel})}\,{v-{u}_x\over V}\boldsymbol{du} \nonumber \\
& = & -2\pi G^2\left(m_f+m\right) \rho\int f(\boldsymbol{u}) {v-u_{x}\over V^3} \ln\left[1+{p_{max}^2 V^4\over G^2 \left(m_f+m\right)^2}\right]\boldsymbol{du} \nonumber
\end{eqnarray}
where $\rho=m_fn$.
If the field star distribution is isotropic
in velocity space, there are various ways to simplify the
integral over velocities.
Chandrasekhar chose one that was especially propitious:
representing the velocity-space volume element in terms of $u$ and $V$ yields 
\begin{eqnarray}
\langle\Delta v_{\parallel}\rangle &=& 
-\frac{16\pi^2 G^2\left(m_f+m\right) \rho}{v^2}\int_0^{\infty} f(u)u^2 
{\cal H}\left(v,u,p_\mathrm{max}\right) du, \nonumber \\
{\cal H}(v,u,p_{max}) &=& {1\over 8u}\int_{|v-u|}^{v+u} dV\left(1 + {v^2-u^2\over V^2}\right) \ln\left[1 + {p^2_{max}V^4\over G^2 (m_f+m)^2}\right].\nonumber \\
\label{Equation:df2}
\end{eqnarray}
This choice separates the integrand into a product of $f$ with a weighting function $\mathcal{H}$ (which Chandrasekhar wrote as $J/8u$).

Chandrasekhar then pointed out an approximation that greatly
simplies $\mathcal{H}$.\footnote{The integral for $\mathcal{H}$ can be 
expressed exactly in terms of simple functions, although the expression
is quite long. This is much easier to demonstrate nowadays than it was
in the 1940s, with tools like Mathematica.}
``Under most conditions of practical interest,'' he wrote,
the argument of the logarithm in equation~(\ref{Equation:df2})
 ``is generally very large compared with unity.''
Invoking this assumption, Chandrasekhar derived various approximate forms for $\mathcal{H}$.
The last, and simplest, of these was
\begin{equation}
{\cal H} = \left\{ \begin{array}{ll}
	\ln\Lambda & \mbox{if $v>u$,} \nonumber \\
	0 & \mbox{if $v<u$}
	\end{array}
	\right.
\end{equation}
which corresponds to treating the logarithmic term in equation~(\ref{Equation:df2})
as a constant.
In this approximation, the dynamical friction coefficient becomes
\begin{equation}\label{Equation:df0}
\langle\Delta v_{\parallel}\rangle = 
-\frac{16\pi^2 G^2 \left(m_f+m\right)\rho\ln\Lambda}{v^2} \int_0^{v} f(u) u^2 du.
\end{equation}
Equation~(\ref{Equation:df0}) contains the beautiful result that the dynamical
friction force is proportional to the number of field stars moving more slowly
(at infinity) than the test star.

For the ``Coulomb logarithm'', Chandrasekhar suggested the form
\begin{equation}
 \ln\Lambda=\ln\left[\frac{p_\mathrm{max}\overline{u^2}}{G(m_f+m)}\right].\nonumber
\end{equation}
As for the parameter $p_\mathrm{max}$, Chandrasekhar and von Neumann 
\cite{ChandrasekharvonNeumann1942}
argued that the effect of gravitational interactions should be cut off at roughly
the interparticle distance.
Later authors \cite[e.g.][]{CohenSpitzerRoutly1950} have generally advocated
larger values for $p_\mathrm{max}$: the physical size of the system,
or the density scale length.

Following the derivation of equation~(\ref{Equation:df0}),
Chandrasekhar wrote: ``As we shall see presently, it is precisely on this account
[i.e. on account of the fact that only stars with $u<v$ contribute to the force]
that dynamical friction appears in our present analysis."

Maybe I missed it, but I could not find where in his paper Chandrasekhar justifies
this statement.
In any case, it is fun to play devil's advocate, 
and ask how dynamical friction acts in cases where this is not true.

A student of Chandrasekhar's, Marvin Lee White, considered one such case
\citep{White1949}.
If $p_\mathrm{max}$ is equated with the half-mass radius of a stellar
cluster, then using the Virial Theorem, it is easy to show that 
$\ln\Lambda\approx\ln(N)$ with $N$ the number of stars in the cluster.
For sufficiently small $N$, the assumption of large $\ln\Lambda$ breaks down.
White showed that in a cluster containing just a few hundred stars,
a significant fraction of the frictional force acting on a test mass
could come from stars with $u>v$.

Two other cases come to mind, both in the context of massive black holes (MBHs).
\begin{itemize}
\item[1] A MBH near the center of a galaxy experiences a kind of
random walk due to perturbations from stars.
Assuming equipartition of kinetic energy, its velocity will be small,
 $v\approx (m_\star/M)^{1/2}\sqrt{\overline{u^2}}$.
But if $v$ is small,
the logarithmic term in equation~(\ref{Equation:df2})  will be close to zero
for all field stars with $u<v$.
Either the frictional force acting on the MBH is much less than implied by equation (\ref{Equation:df0}),
or most of the  force must come from field stars with $u>v$.
\item[2] There are physically reasonable models for the distribution 
of stars around a MBH that imply few or no stars with velocities less than
some value; for instance, the local circular velocity.
In such cases, Chandrasekhar's approximate formula would predict  no frictional 
force on a body moving in a circular orbit; all 
the force would have to come from the fast-moving stars.
\end{itemize}

Consider first the Brownian motion example.
Chandrasekhar (1943) showed that the rms velocity of the test star
is given by
\begin{equation}
v^2_\mathrm{rms} = \frac32\left|\frac{\langle(\Delta v_\parallel)^2\rangle}
{\langle\Delta v_\parallel\rangle/v}\right|.\nonumber
\end{equation}
Expanding the diffusion coefficients about $v=0$,
\begin{eqnarray}
\langle\Delta v_\parallel\rangle &=& -Av+ Bv^3\ldots , \label{expand1} \nonumber\\
\langle(\Delta v_\parallel)^2\rangle &=& C + Dv^2\ldots  \label{expand2} \nonumber
\end{eqnarray}
so that $v^2_\mathrm{rms}=3C/2A$.
The coefficients $A$ and $C$ are obtained from the low-velocity limit
of equation~(\ref{Equation:df2}),
and the equivalent expression for $\langle(\Delta v_\parallel)^2\rangle$ \cite{Chandrasekhar1942}.
The results are
\begin{subequations}
\begin{eqnarray}
A &=& {32\over 3}\pi^2 G^2 M\rho \int_0^{\infty}
f(u) {p_{max}^2u^4/G^2M^2\over 1+p_{max}^2u^4/G^2M^2} {du\over u} , 
\label{Equation:A} \\
C &=&  {16\over 3}\pi^2G^2m_\star\rho \int_0^\infty f(u)
\log\left(1+{p_{max}^2u^4\over G^2M^2}\right)u\, du\label{Equation:C}
\end{eqnarray}
\end{subequations}
where $M$ is the MBH mass.
It is evident from the first of these expressions that field stars of every velocity contribute
to the dynamical friction force.
% acting on the massive body.
If the field star velocity distribution is a Maxwellian,
the result of the integrations is
\begin{equation}
A =  4\sqrt{6\pi} G^2M \rho\left(\overline{u^2}\right)^{-3/2} \ln\Lambda^\prime, \ \ \ \ \ 
C = \frac83\sqrt{6\pi} G^2m_\star\rho\left(\overline{u^2}\right)^{-1} \ln\Lambda^\prime
\nonumber
\end{equation}
where
\begin{equation}
\ln\Lambda'  \equiv  {1\over 2} \int_0^{\infty} dz\ e^{-z} \ln\left(1+ 
\frac{4p_\mathrm{max}^2(\overline{u^2})^2}{9G^2M^2}z^2\right) \approx \ln\sqrt{1 + {2p^2_{max}(\overline{u^2})^2\over 9G^2M^2}}. \nonumber
\end{equation}
The rms velocity of the MBH has the expected value,
$(m_\star\overline{u^2}/M)^{1/2}$,
regardless of the value of $\ln\Lambda'$ .
But returning to the question of {\it which} stars generate the frictional 
force: the integrand of equation~(\ref{Equation:A}) peaks at
$u\approx (GM/p_\mathrm{max})^{1/2}$.
Furthermore, $p_\mathrm{max}$ in the case of a MBH near the center of a galaxy is
expected to be comparable to the MBH influence radius $GM/\overline{u^2}$ \cite{Maoz1993,MM2001}.
It follows that most of the frictional force responsible for Brownian motion of a MBH
comes from field stars with speeds similar to their rms value, and not exclusively,
or even predominantly,
from the slow-moving stars (Figure~1).

\begin{figure}[ht]
\begin{center}
\includegraphics[width=0.60\columnwidth,angle=90.]{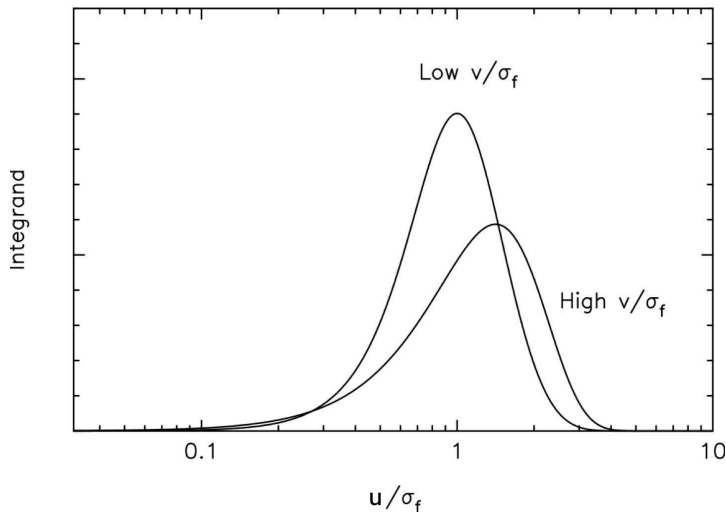}
\caption{Relative contribution of different field-star velocities to the 
dynamical friction integral, equation~(\ref{Equation:A}),
in the limit of low and high velocities of the
test mass, assuming
a Maxwellian distribution of field-star velocities with one-dimensional
velocity dispersion $\sigma_f=(\overline{u^2}/3)^{1/2}$.
$p_\mathrm{max}=GM/\sigma_f^2$ has been assumed, appropriate for
a massive test body moving at the center of a galaxy.
Roughly the same set of field stars dominates the frictional force,
whether the test mass is moving very slowly or very fast.}
\label{Figure:LowHighV}
\end{center}
\end{figure} 

The second case mentioned above was a test mass orbiting around a 
MBH at the center of a galaxy.
Suppose that the stars, which are responsible for the frictional force,
are distributed around the MBH
with density $\rho(r)\propto r^{-\gamma}$.
Assuming an isotropic velocity distribution, their phase space density is given by
\begin{equation}\label{Equation:fofv}
f \propto (-E_f)^{\gamma-3/2}\propto \left(2v_c^2-u^2\right)^{\gamma-3/2}
\end{equation}
where $v_c=(GM/r)^{1/2}$ is the local circular speed.
%The phase space density is zero for $u\ge v_\mathrm{esc}=2^{1/2}v_c$,
%the escape velocity from the MBH.
For $\gamma<3/2$, $f$  diverges at $u=2^{1/2}v_c$,
and in the limit $\gamma=1/2$, $f$ is a delta-function at $u=2^{1/2}v_c$.
In other words, in a $\rho\propto r^{-1/2}$ density cusp around a MBH,
{\it none} of the stars at $r$ is moving more slowly than the local circular 
speed.
The dynamical friction formula in its approximate form~(\ref{Equation:df0}) 
would predict zero frictional force in this case.

It so happens that the distribution of
stars at the centers of galaxies containing MBHs is often as flat as
$\rho\sim r^{-1/2}$ \cite{Cote2007}; 
indeed this appears to be the case at the center of the Milky Way
\citep{Buchholz2009}.
One explanation is that these low-density cores were ``carved out'' by binary
MBHs \cite{Graham2004}.\footnote{If so, then one might expect $f$ to be somewhat {\it an}isotropic.}
But whatever their origin, the argument just given suggests that dynamical friction in such cores must be due almost entirely to stars that move faster than the inspiralling body.
Of course, the frictional force on a test mass at a distance $r$ from 
the MBH would come partly from stars at greater $r$, and some of these stars will have 
smaller $u$. 
Chandrasekhar's theory does not make clear predictions in the
case of inhomogeneous systems.
It would be interesting to address 
this question using an $N$-body code.

Why do the two populations -- stars with $u<v$ and stars with $u>v$ --
contribute in such different ways to the frictional force?
One way to visualize this is to compute the dynamical-friction ``wake,''
the overdensity of stars behind the test body that is responsible for the
decelerating force.
If the field star distribution is infinite and homogeneous, and if the
test body's velocity is rectilinear and unaccelerated (all assumptions
that were made by Chandrasekhar), the wake is time-independent in a frame
following the test mass, and its density
can be computed using Jeans's theorem given the distribution of field star
velocities at infinity \cite{Mulder1983}.

Figure~2 shows the result of such a calculation, assuming that the velocity distribution at
infinity is given by equation~(\ref{Equation:fofv}) with $\gamma=5/4$ 
and that the test body's velocity is $v_c$.
For this choice of $f$, the ``fast'' stars dominate both 
the total density at infinity, as well as the density in the wake.
\begin{figure}
\begin{center}
  \includegraphics[angle=90.,width=4.5in]{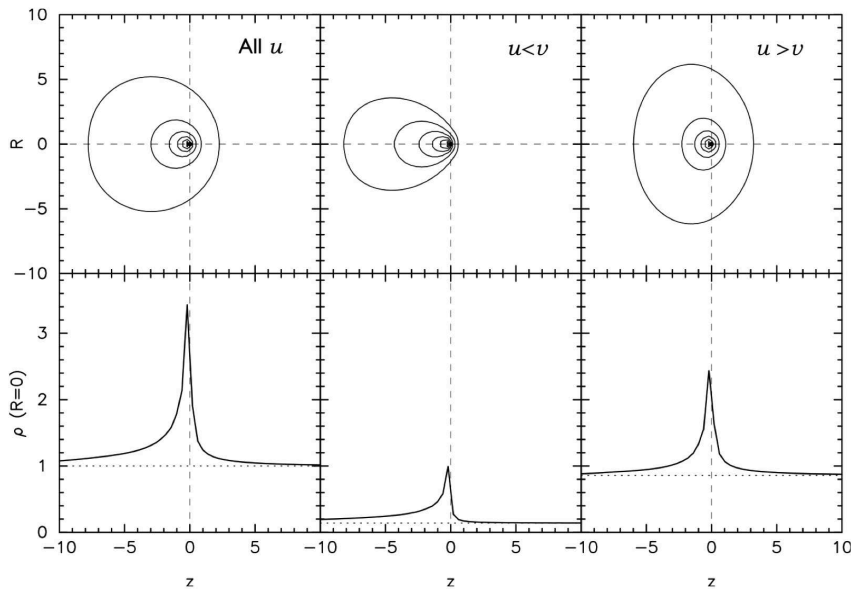}
  \caption{Dynamical friction wake around a massive object moving
through an infinite homogeneous stellar system.
The panels at the top show contours of the density, in a frame that
follows the massive body (located at the origin); motion is to the
right at constant speed $v$.
The left panel shows the total density, the middle panel shows the density 
contributed only by stars with velocity at infinity less than that 
of the massive body, and the right panel shows the contribution
from the complement of stars that move faster than $v$.
The lower panels show the density along a line through
the moving body in the direction of its motion.
In the usual approximation, first derived by Chandrasekhar, one computes
the dynamical friction force ignoring the fast-moving stars.
This figure helps to explain why that is a good approximation,
even if the fast stars dominate the total density.}
\label{fig:mulder}
\end{center}
\end{figure}
However the shapes of the two, partial density wakes are very different.
The wake created by the fast stars is elongated counter to the direction 
of the test body's motion, reducing the net force that it exerts on the
test mass. 
In addition, the change in density between the upstream and downstream
sides of the test mass is less for the fast stars than for the slow stars.
These two differences are responsible for the small contribution of the
fast stars to the total frictional force, in spite of their higher density
at infinity and in the wake.

\section{Relativistic Dynamical Friction}

In 1969, Edward E. Lee, a PhD student of Chandrasekhar's, re-derived the
expression for the dynamical friction force, using the first post-Newtonian
(1PN) approximation for the relative orbit of test and field stars \cite{Lee1969}.
Lee motivated his calculation in the following way: 
``These effects of general relativity, in the lowest order, may be relevant for
consideration of stellar encounters in dense systems such as are now contemplated 
in a number of contexts.''
% (see Gold, Axford, and Ray 1965; Thorne and Ipser 1968).''
Lee's final expression for the relativistic dynamical friction coefficient
was fairly complex, and he did not go so far as to evaluate it in the case of
a particular velocity distribution.
But given $f(u)$, Lee noted that the relativistic formula predicted a higher frictional
force than the non-relativistic formula, due to the greater relative 
deflection in the 1PN approximation, and also because the relativistic
transformation from the inertial frame to the frame of the test mass
introduces an additional term into the force.

\section{Relativistic Stellar Dynamics}

Massive black holes were not yet quite in vogue at the time that
Lee wrote his thesis.
It is now commonly assumed that all massive galaxies have them.
Given Chandrasekhar's fundamental contributions to the theory of 
gravitational encounters, 
and to the theory of black holes \cite{Chandrasekhar1983},
gravitational encounters between stars orbiting near a MBH is a topic
that he might naturally have been interested in.

A black hole is a compact object, and from the point of view of a star
 that orbits far outside the event horizon, its gravitational field should be nearly indistinguishable from that of a Newtonian point mass.
But the Newtonian approximation breaks down for matter
that passes within a few gravitational radii:
\begin{equation}\label{Equation:DefineRG}
r_g \equiv\frac{GM}{c^2}
\approx 5\times 10^{-8} \mathrm{pc}\left(\frac{M}{10^6M_\odot}\right) \nonumber
\end{equation}
since the orbital velocity at such distances approaches the speed of light.

At first sight, one is struck by the enormous difference between
$r_g$ and $r_\mathrm{infl}$, the radius of gravitational of
influence of a MBH:
\begin{equation}
r_\mathrm{infl} \equiv \frac{GM}{\sigma_\star^2}
\approx 1 \mathrm{pc} \left(\frac{M}{10^6 M_\odot}\right)
\left(\frac{\sigma_\star}{100 \mathrm{km\ s}^{-1}}\right)^{-2}\nonumber
\end{equation}
where $\sigma_\star$ is the stellar velocity dispersion at $r\gtrsim r_\mathrm{infl}$.
These two equations suggest that the vast majority of stars within the
influence sphere are too far from the MBH\ for relativity to be
important.

This conclusion turns out to be misleading, for a couple of reasons.
First: the effects of relativity depend less on the size of an orbit than on
its distance of closest approach to the MBH. 
The lowest-order corrections to the Newtonian
equations of motion have amplitudes that are of order ${\cal P}^{-1}$
where ${\cal P}$ is the 
{\it penetration parameter}:
\begin{equation}
{\cal P} \equiv (1-e^2)\frac{a}{r_g} = \left(1+e\right)\frac{r_p}{r_g}. \nonumber
\end{equation}
Here, $a$ and $e$ are the semi-major axis and eccentricity of the orbit,
and $r_p=(1-e)a$ is the radius of periapsis.
It turns out that the feeding of stars and compact objects to MBHs\ occurs
predominantly from very eccentric orbits.
For instance, capture of stellar-mass BHs by MBHs\ -- 
so-called ``extreme-mass-ratio inspiral''
(EMRI) -- is believed to take place from orbits with semi-major axes
$a\lesssim 0.1$ pc and eccentricities in the range $0.999-0.99999$
\cite{HA2006}.
For such orbits, ${\cal P}$ can be of order unity even though the orbit extends
outward to thousands or tens of thousands of gravitational radii.

A second reason why the effects of relativity can not be ignored has
to do with the way in which stars get placed onto orbits of such high
eccentricity.
The dominant mechanism is believed to be torques, i.e. non-radial
forces, that arise from the slightly aspherical distribution of matter 
near a MBH \citep{RT1996}.
These torques remain effective as long as orbits near the MBH
-- both the orbit of the star being torqued, and the orbits of the 
torquing stars -- 
maintain their orientations;
any mechanism that causes orbits to precess (for instance) tends to 
randomize the torques.
Relativistic precession of the periapsis -- or, as it is know in the
Solar system, precession of the perihelion -- is such a mechanism.
If the time scale for relativistic
precession is shorter than the time time required for the torques to do
their work, feeding of objects to the MBH\ will be greatly inhibited.
This relativistic quenching effect turns out to be of major importance
in the EMRI problem.

Consider a star on a bound orbit near a MBH, with semi-major
axis $a$ and angular momentum $\boldsymbol{L}$.
The orbit-averaged torque $|\boldsymbol{F}\times\boldsymbol{r}|$
that another star of mass $m$ exerts on it is $\sim Gm/a$.
The residual torque due to $N$ stars, randomly oriented about the MBH at 
about the same radius, is $\left|\boldsymbol{T}\right|\approx\sqrt{N}Gm/a$. 
Over some span of time, this orbit-averaged torque is
nearly constant, and the angular momentum 
of a test star's orbit responds by changing linearly with time,
$\Delta L/\Delta t\approx T\approx \sqrt{N}Gm/a$.
This coherent evolution continues for a time $\sim t_\mathrm{coh}$,
where $t_\mathrm{coh}$ is the time scale associated with the most rapid
mechanism that randomizes the torques, i.e. the orientations of the
orbits resposible for the torques.

Sufficiently far from a MBH, coherence breaking is due mostly to the same
distributed mass that generates the torques.
Modelling that mass as spherical,
the associated precession time is
\begin{equation}\label{eq:tm}
t_M\approx P\frac{M}{M_\star}
%\left(1-e^2\right)^{-1/2} 
\end{equation}
where $P$ is the orbital period, $M$ is the MBH mass, and $M_\star$ is the distributed mass
within $r=a$.
(There is also a dependence on $e$, in the sense that
eccentric orbits precess more slowly.)
This time is believed to be of order $10^4$ yr in the case of the 
so-called S stars, bright young stars at the center of the Milky Way
whose orbits can be tracked astrometrically; the S stars have 
$0.01\ \mathrm{pc}\lesssim a\lesssim 0.1\ \mathrm{pc}$ and
 $10^1\mathrm{yr}\lesssim P\lesssim 10^3\mathrm{yr}$
\citep{Gillessen2009}.

Closer to the MBH, relativistic effects become important.
The most important precessional time scales associated with
relativistic corrections to the equations of motion are
\begin{subequations}
\begin{eqnarray}
t_S &=& \frac16 {\cal P} P \nonumber \\
t_J &=& \frac14 {\cal P}^{3/2}\chi^{-1} P \nonumber \\
t_Q &=& \frac13 {\cal P}^2 \chi^{-2} P\nonumber 
\end{eqnarray}
\end{subequations}
The first of these refers to precession of the periapse; the subscript $S$
indicates that this effect is due to the Schwarzschild (zero spin)  part of the metric.
This ``Schwarzschild precession'' is similar to, but in the opposite sense of, 
precession due to the distributed mass.
The other two time scales are associated with the spin ($J$) and quadrupole
moment ($Q$) of the MBH. 
$t_J$ is the time for precession of the line of nodes due to frame-dragging,
and $t_Q$ is the nodal precession time resulting from the nonsphericity of space-time
around a spinning hole.
(The quadrupole precession rate also depends on the inclination of the
orbit with respect to the MBH spin axis.)
These latter two times are functions of $\chi$, the dimensionless spin of the MBH:
writing the hole's spin angular momentum as $\boldsymbol{J}$,
\begin{equation}
\boldsymbol{\chi} \equiv \boldsymbol{J}\frac{GM^2}{c^2}\nonumber
\end{equation}
and $\chi=1$ for a maximally-spinning hole. 

The time scales for relativistic precession defined above decrease as $(1-e)$, or
faster, as $e\rightarrow 1$.
If torques from the $\sqrt{N}$ asymmetries in the stellar distribution
drive the eccentricity of a test star's orbit to a sufficiently large value, 
relativistic effects will 
dominate its precession.
At some critical precession rate (i.e. eccentricity),
the sign of the torque as experienced by the orbit will fluctuate
with such a high frequency that its net effect over one precessional period 
will be negligible:
in other words, relativity will ``quench'' the effects of the torques.

This critical eccentricity can be estimated as follows.
As shown above, the residual torque produced by $N$ stars is
\begin{equation}\label{eq:RRtorque}
T\approx \frac{Gm}{a} \sqrt{N(a)}
\approx \frac{1}{\sqrt{N(a)}}\frac{GM_\star(a)}{a}\nonumber
\end{equation}
where $N(a)=M_\star(a)/m$ is the number of stars, of mass $m$, within radius $r=a$.
Writing $L=\left[GM a(1-e^2)\right]^{1/2}$ for the test star's
angular momentum, 
the time scale over which $L$ is changed by this torque is
\begin{equation}
\left|\frac{1}{L}\frac{dL}{dt}\right|^{-1} \approx 
\sqrt{N(a)}\frac{M}{M_\star(a)}
\left[\frac{a^3(1-e^2)}{GM}\right]^{1/2}.\nonumber
\end{equation}
The condition that this time be shorter than the relativistic
precession time $t_S$ is 
\begin{equation}
\ell> \ell_\mathrm{SB}\approx 
\frac{r_g}{a} \frac{M}{M_\star(a)}\sqrt{N(a)}\nonumber
\end{equation}
where 
\begin{equation}
\ell\equiv L/L_c=(1-e^2)^{1/2}\nonumber
\end{equation}
is the ratio of the test star's angular momentum to its circular value.
In terms of eccentricity,
\begin{eqnarray}\label{eq:SBd}
\left(1-e^2\right)_\mathrm{SB} &\approx&
2\left(\frac{r_g}{a}\right)^2 
\left(\frac{M}{m}\right)^2 
\frac{1}{N} \\
&\approx& 2\times 10^{-5}
\left(\frac{a}{0.01\mathrm{pc}}\right)^{-2}
\left(\frac{M}{10^6M_\odot}\right)^{4}
\left(\frac{m}{10M_\odot}\right)^{-2}
\left(\frac{N}{10^2}\right)^{-1}. \nonumber
\end{eqnarray}
This ``Schwarzschild barrier'' (so called since it arises from
the Schwarzschild part of the metric) sets an effective upper limit to
the eccentricty of the test star. 

\begin{figure}
\begin{center}
  \includegraphics[angle=90.,width=4.0in]{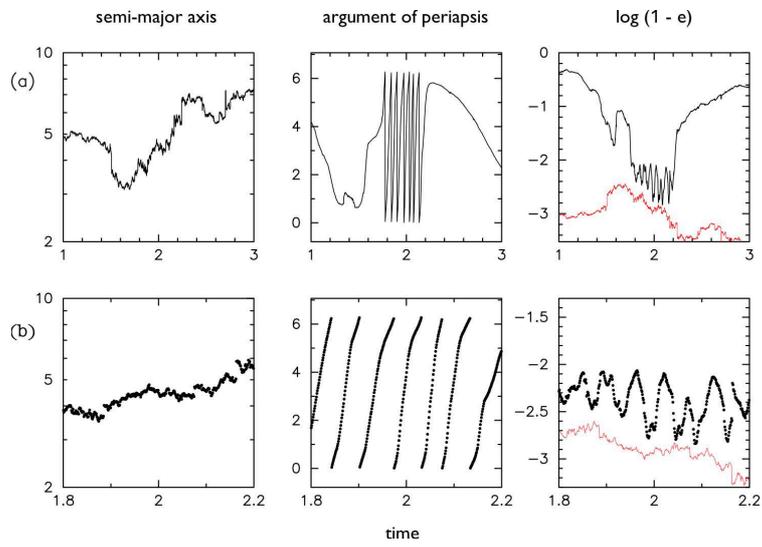}
  \caption{An eccentric orbit near a massive black hole,
extracted from an $N$-body integration \cite{MAMW2011}.
Plotted are the semi-major axis, argument of periapse and eccentricity versus time,
at low (top) and high (bottom) time resolutions.
In the plots of eccentricity vs. time, the lower (red) curves
show the predicted location of the Schwarzschild  barrier.
Changes in the predicted barrier location reflect changes in the
semi-major axis, via eq.~(\ref{eq:SBd}).}
\label{fig:Nbody}
\end{center}
\end{figure}

What happens when a star (or stellar remnant) ``strikes'' the barrier?
Figure~\ref{fig:Nbody} shows one example, extracted from an $N$-body
simulation \cite{MAMW2011}.
The orbit's eccentricity first oscillates, about a lower bound given
by equation~(\ref{eq:SBd}).
The oscillations have a period equal to
$t_S$; they reflect the periodically changing effects of the torques on 
the precessing star.
After some elapsed time -- roughly 6 precessional periods in this case -- 
the star is ``reflected'' from the barrier by the torques, back to smaller values of the
eccentricity.
The elapsed time until reflection is just the coherence time 
$t_\mathrm{coh}$ defined above.
The orbits of all the other stars are also precessing: not as rapidly as the
test star (which has the highest eccentricity), but at a lower rate --
in the example shown in Figure~\ref{fig:Nbody}, the distributed mass
determines the precession rate of most stars, via equation~(\ref{eq:tm}).
After a time of $t_\mathrm{coh}\approx t_\mathrm{M}$, 
the net torque due to the other stars has changed its direction 
and the angular momentum of the test star responds by decreasing,
taking the star away from the barrier.

The $N$-body experiments from which Figure~\ref{fig:Nbody} was derived
were designed to generate EMRIs, i.e. inspiral of stellar remnants into the MBH.
For this to happen, the remnants must sometimes penetrate the
Schwarzschild barrier.

Penetration does occur, but because the coherent torquing -- or
``resonant relaxation'' -- is quenched by the effects of relativity, 
all that is left is the kind of random gravitational scattering treated by Chandrasekhar,
or ``non-resonant relaxation''.
It turns out that the rate at which objects get scattered past the
angular momentum barrier
%onto capture orbits (orbits that are able to 
%shrink via gravitational wave emission before being scattered away), 
is accurately predicted by Chandrasekhar's formulae.
The only additional consideration arises from the fact that stars 
remain near the barrier only for a limited time, $\Delta t\approx t_\mathrm{coh}$,
before the changing background torques pull them away;
to be effective, the encounters described by Chandrasekhar's formulae 
must decrease the test body's angular momentum,
by an amount greater than the amplitude of the oscillations shown
in Figure~\ref{fig:Nbody}, in a time less than $t_\mathrm{coh}$.
This argument correctly reproduces the EMRI capture rate observed
in the simulations, and also turns out to imply a minimum semi-major axis, of
order $10^{-3}$ pc, below which stars can not be driven past the 
barrier, either by resonant or non-resonant relaxation \cite{MAMW2011}.

\bigskip
The collisional dynamics of relativistic stellar systems is 
a fascinatingly rich topic.
It is a shame that Chandrasekhar is not here to sort it out for us.

\acknowledgments
I thank Richard Miller, Alar Toomre and Peter Vandervoort for sharing with me their reminiscences of Chandrasekhar and his work on stellar dynamics.

\bibliography{references}

\end{document}